\documentclass{elsart3}

\usepackage{epsfig}

\newcommand{\cf}[1]{\langle #1 \rangle}                      
\newcommand{\phdagger}{\mathop{\phantom{\dagger}}}           
\newcommand{\psiop}[1]{\psi^{\phdagger}_{#1}}                
\newcommand{\psidop}[1]{\psi^{\dagger}_{#1}}                 
\newcommand{\bop}[1]{b^{\phdagger}_{#1}}                     
\newcommand{\bdop}[1]{b^{\dagger}_{#1}}                      
\newcommand{\fop}[1]{f^{\phdagger}_{#1}}                     
\newcommand{\fdop}[1]{f^{\dagger}_{#1}}                      

\newcommand{\bml}{\begin{mathletters}}
\newcommand{\eml}{\end{mathletters} \hspace{-5pt}}

\usepackage{amssymb}

\begin{document}

\begin{frontmatter}



\title{Conformal Field Theory Approach to the \\ Two-Channel Anderson Model}

\author[Goteborg]{H. Johannesson\corauthref{cor1}},
\author[Rutgers]{ N. Andrei}, and
\author[Geneva]{C. J. Bolech}
\address[Goteborg]{Theoretical Physics, Chalmers and G\"oteborg University,
G\"oteborg, Sweden}
\address[Rutgers]{Center for Materials Theory, Rutgers University, USA}
\address[Geneva]{School of Physics, University of Geneva, Switzerland}
\corauth[cor1]{Corresponding author. \\ {\em E-mail:
johannesson@fy.chalmers.se}}


\begin{abstract}
The two-channel Anderson impurity model serves as a prototype for describing heavy-fermion materials with a possible
mixed-valent regime with both quadrupolar and magnetic character. We report on the low-energy physics of the model, using
a conformal field theory approach with exact Bethe Ansatz results as input.
\end{abstract}

\end{frontmatter}


As is well-known, several uranium-based heavy fermion materials exhibit manifest non-Fermi liquid behavior
at low temperatures. A particularly intriguing case is that of the UBe$_{13}$ compound. It has been suggested
that the electric quadrupolar degrees of freedom of the uranium ions (in 5$f^2$ configurations) in the Be$_{13}$
host are overscreened by the local orbital motion of the conduction electrons, producing a two-channel Kondo
response of this material \cite{Cox}. Taking into account the possibility that the magnetic 5$f^3$ configurations
may also be at play, one is led to consider the two-channel Anderson impurity model \cite{Cox}
\begin{eqnarray}
\ \ \ \ \ \ \ \ H &=& H_0
+ \varepsilon_s \fdop{\sigma} \fop{\sigma} +
\varepsilon_q \bdop{\bar{\alpha}} \bop{\bar{\alpha}} \nonumber \\
\ \ \ \ \ \ \ \  &+&V (\psidop{\alpha\sigma}(0) \bdop{\bar{\alpha}} \fop{\sigma} +
\fdop{\sigma} \bop{\bar{\alpha}} \psiop{\alpha\sigma}(0)) \, ,
\label{model}
\end{eqnarray}
where $H_0$ is the free part of the Hamiltonian.
The conduction electrons $\psidop{\alpha\sigma}$ carry
spin $(\sigma = \uparrow, \downarrow)$ {\em and} quadrupolar $(\alpha = \pm)$ quantum numbers,
and hybridize with a local uranium ion via a matrix element $V$.
The ion is modeled by a quadrupolar
[magnetic] doublet of energy $\epsilon_q$ [$\epsilon_s$], created by a boson [fermion] operator
$\bdop{\alpha}$ [$\fdop{\sigma}$]. Strong Coulomb repulsion implies single occupancy of the
localized levels: $\fdop{\sigma} \fop{\sigma} + \bdop{\bar{\alpha}} \bop{\bar{\alpha}} = 1$.

We have carried out a nonperturbative analysis of the model, exploiting boundary conformal field theory \cite{Affleck}
to trade the hybridization interaction in (\ref{model}) for a scale invariant boundary condition on the conduction
electrons. Identifying the proper boundary condition immediately identifies the critical theory to which the model belongs.
By a numerical fit to the exact Bethe Ansatz solution \cite{Bolech} of (\ref{model}) we can determine {\em all} parameters
and scales of this theory, thus obtaining a complete description of the low-energy dynamics of the model.
Our most important results can be summarized:

The model renormalizes to a line of fixed point Hamiltonians parameterized by the average charge $n_c \equiv \cf{\fdop{\sigma}
\fop{\sigma}}$ at the impurity site (Fig. 1), all exhibiting the same zero-temperature impurity entropy $S_{imp} = k_B$ln$\sqrt{2}$,
typical of two-channel Kondo physics \cite{Johannesson}.
The low-temperature specific heat induced by the impurity takes the form $C_{imp} = \mu_s T$ln$(T_s/T) +
\mu_q T$ln$(T_q/T) + O(T)$, where $\mu_{s\, [q]}$ and $T_{s\, [q]}$ are amplitudes and temperature scales respectively,
measuring the participation of the spin [quadrupolar] degrees of freedom in the screening process. Similarly, the impurity
response to a magnetic [quadrupolar] field is also of two-channel Kondo type, $\chi^{s\, [q]}_{imp} \sim \mu_{s\, [q]}
$ln$(T_{s\, [q]}/T)$, with the amplitudes $\mu_{s\, [q]}$ decreasing exponentially with $\epsilon$ [$-\epsilon$].
We have also calculated the Fermi edge singularities
caused by {\em time-dependent} hybridization between conduction electrons and impurity. These are coded by the scaling dimensions
of the pseudo-particles $\bdop{\bar{\alpha}}$ and $\fdop{\sigma}$ in (\ref{model}), and we find that $x_b = (3+2n_c^2)/16$ and
$x_f = (5-4n_c +2n_c^2)/16$ respectively.
We point out that the values of the pseudo-particle scaling dimensions as obtained by various approximation schemes,
e.g. NCA and large-$N$ calculations \cite{Ruckenstein}, deviate from our exact result.

The conformal field theory formalism allows us to determine also the asymptotic dynamical properties of the model,
incl. Green's functions and resistivities. We have calculated the self-energy of the electrons
and find that it leads to a zero-temperature resistivity {\em independent} of the level separation $\epsilon$:
$\rho(T\!=\!0) = 3n_i/4\pi (e\nu v_F)^2 $, with $n_i$ the (dilute) impurity concentration, $\nu$ the density of states at the
Fermi level, and $v_F$ the Fermi velocity. The finite-temperature calculation is in progress.
\\ \\ \\ \\ \\

\begin{figure}
\begin{center}
\includegraphics[width=0.4\textwidth]{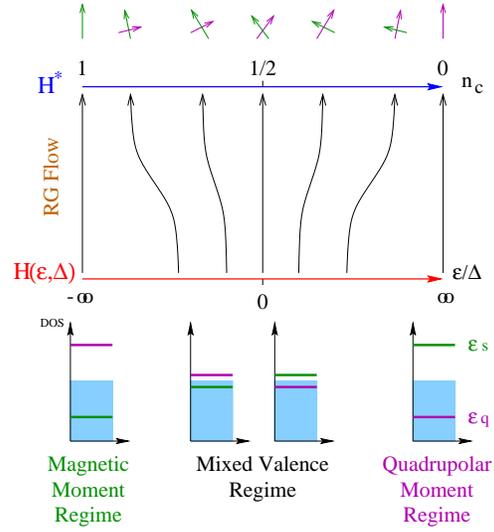}
\caption[Schematic RG Flow.]{Schematic renormalization group flow diagram of
  the two-channel Anderson impurity model. The flow connects the line of
  microscopic theories characterized by the ratio $\varepsilon/\Delta$ ($\epsilon =
  \epsilon_s - \epsilon_q, \,\Delta = \pi \rho V^2$) with
  the line of infrared fixed points parameterized by $n_c$.}
\label{Aflow}
\end{center}
\end{figure}

\subsection*{ }
\subsection*{ }




\begin{thebibliography}{9}

\bibitem{Cox}For a review, see D. L. Cox and A. Zawadowski, Adv. Phys. {\bf 47}, 599 (1998).

\bibitem{Affleck} I. Affleck and A. W. W. Ludwig, Nucl. Phys. B {\bf 360}, 641 (1991).

\bibitem{Bolech} C. J. Bolech and N. Andrei, Phys. Rev. Lett. {\bf 88}, 237206 (2002).

\bibitem{Johannesson} H. Johannesson, N. Andrei, and C. J. Bolech, Phys Rev. B {\bf 68}, 075112 (2003).

\bibitem{Ruckenstein} D. L. Cox and A. E. Ruckenstein, Phys. Rev. Lett.
{\bf 71}, 1613 (1993).




\end{thebibliography}
\end{document}